\documentclass[12pt,preprint]{aastex}
\usepackage{epsfig}
\usepackage{natbib}
\usepackage{amssymb, amsmath, amsbsy, epsfig, epsf, slashbox, rotating}
\usepackage[dvips]{color}

\def\A{${\rm \AA}$}
\newcommand{\lya}{Ly$\alpha$}
\newcommand{\lyb}{Ly$\beta$}
\newcommand{\feii}{Fe\,{\footnotesize II}}

\newcommand{\mgii}{Mg\,{\footnotesize II}}
\newcommand{\heii}{He\,{\footnotesize II}}
\newcommand{\nv}{N\,{\footnotesize V}}
\newcommand{\niv}{N\,{\footnotesize IV}}
\newcommand{\niii}{N\,{\footnotesize III}]}
\newcommand{\oi}{O\,{\footnotesize I}}
\newcommand{\ovi}{O\,{\footnotesize VI}}
\newcommand{\oiii}{O\,{\footnotesize III}]}
\newcommand{\oiv}{O\,{\footnotesize IV}]}
\newcommand{\cii}{C\,{\footnotesize II}}
\newcommand{\ctwo}{C\,{\footnotesize II}]}
\newcommand{\ciii}{C\,{\footnotesize III]}}
\newcommand{\civ}{C\,{\footnotesize IV}}
\newcommand{\aliii}{Al\,{\footnotesize III}}
\newcommand{\siii}{Si\,{\footnotesize II}}
\newcommand{\siiii}{Si\,{\footnotesize III]}}
\newcommand{\siiv}{Si\,{\footnotesize IV}}
\newcommand{\neiii}{[Ne\,{\footnotesize III}]}

\righthead{Metallicity and Outflow} \shorttitle{Metallicity and
Outflow}
\shortauthors{Wang H.Y. et al.} %\received{......}

\begin{document}
%\begin{CJK*}{GBK}{com}
\title {Metallicity and Quasar Outflows}
\author{Huiyuan Wang\altaffilmark{1,2}, Hongyan Zhou\altaffilmark{1,2,3}, Weimin
Yuan\altaffilmark{4} and Tinggui Wang\altaffilmark{1,2}}
\altaffiltext{1}{Key Laboratory for Research in Galaxies and
Cosmology, University of Science and Technology of China, Chinese
Academy of Sciences, Hefei, Anhui 230026, China;
whywang@mail.ustc.edu.cn} \altaffiltext{2}{Department of
Astronomy, University of Science and Technology of China, Hefei,
Anhui 230026, China}\altaffiltext{3}{Polar Research Institute of
China, Jinqiao Rd. 451, Shanghai, 200136,
China}\altaffiltext{4}{National Astronomical Observatories,
Chinese Academy of Sciences, Beijing 100012, China}

\begin{abstract}
Correlations are investigated of the outflow strength of quasars,
as measured by the blueshift and asymmetry index (BAI) of the
\civ\ line (Wang et al. 2011), with intensities and ratios of broad emission lines,
based on composite quasar spectra built from the Sloan Digital Sky
Survey. We find that most of the line ratios of other ions to
\civ\ prominently increases with BAI. These behaviors can be well understood in the context
of increasing metallicity with BAI. The strength of dominant coolant, \civ\ line, decreases and
weak collisionally excited lines increase with gas metallicity as a result of the
competition between different line coolants. Using \siiv$+$\oiv/\civ\ as an indicator of gas metallicity, we present,
for the first time, a strong correlation between the metallicitiy and the outflow strength of quasars over a wide range of
1.7 to 6.9 times solar abundance. Our result implies that the metallicity plays an important role in the formation of
quasar outflows, likely via affecting outflow acceleration. This effect may have a profound
impact on galaxy evolution via momentum feedback and chemical enrichment.
\end{abstract}


\section{Introduction}

Outflows appear to be a common phenomenon in quasars. On small
scale, mass loss due to outflows is an important component in the
overall structure of quasars (Crenshaw et al. 2003, and references
therein). Recently, outflows have been regarded as one of the main
components in the unification of various subclasses of quasars
(Richards et al. 2011). On large scales, outflows are considered
to be able to clear out inter-stellar medium (ISM; Silk \& Rees
1998; Fabian 1999; Di Matteo et al. 2005) and enrich the
inter-galactic medium (IGM; Friaca \& Terlevich 1998), and
consequently affect the star formation in the host galaxies.
Quasar outflows are important for our understanding of the
co-evolution of galaxies and their central black holes.

There exist, in quasar spectra, prominent features associated with
outflows, e.g. broad absorption lines (BALs) and blueshifted broad
emission lines (BELs). Studies on the two different features
reveal quite similar dependencies of outflow properties on some
fundamental quantities. BAL quasars with larger outflow velocity
tend to have higher Eddington ratio and weaker intrinsic X-ray
relative to UV (Ganguly et al. 2007; Fan et al. 2009). And the
fraction of BAL quasars also increases with Eddington ratio
(Ganguly et al. 2007; Zhang et al. 2010). The blueshift of \civ\
BELs is more strongly correlated with the Eddington ratio than the
BAL properties (Wang et al. 2011, hereafter Wang11) and also
depend on the shape of the ionizing continuum (Leighly \& Moore
2004; Richards et al. 2011). Furthermore, the blueshift is
apparently stronger in radio-quiet quasars than in radio-loud ones
(Marziani et al. 1996). These observational results provide strong
support for the scenario that the outflow is directly related to
the accretion process and driven by radiation pressure.

Is the launch of outflows affected by other factors, in particular
the environment where quasars reside in? In this letter, we
uncover a strong correlation between the outflow strength and
the quasar metallicity. This correlation is actually not unexpected considering
that high metallicity can increase the acceleration of outflows.
One good example with strong outflows and enriched metallicity is
the ultraluminous infrared quasar Q1321$+$058 (Wang et al. 2009).
Since metals are produced by star formation process, this
correlation strongly suggests that star formation in galaxies can
affect the evolution of the central black holes. The connection
between galaxies and black holes might be more intimate and
complex than previous expectation.

This letter is organized as follows. In Section \ref{secsam} we
show the sample selection and present composite spectra as a
function of the blueshift and asymmetry index (BAI) of the \civ\
line, which has been shown to be a good indicator of the strength
of quasar outflows (Wang11). We also measure the line flux from
the composites. In Section \ref{sec_bmc} we present strong
correlations of various line flux ratios with BAI. These
results provide strong evidence to support that outflow strength
increases with the metallicity. Finally, we summarize the results
and discuss their implications in Section \ref{secsum}.

\section{Sample Selection and Composite Spectra}\label{secsam}

\subsection{Sample Selection}\label{sec_ss}

We select quasars in the redshift range $1.7<z<4.0$ from the Fifth
Data Release of the Sloan Digital Sky Survey (SDSS; Schneider et
al. 2007). The redshift range is chosen in such a way that the
\civ\ line falls in the wavelength coverage of the SDSS
spectrograph. To ensure reliable measurements of emission line
parameters, we select objects with median signal-to-noise ratio
(S/N)$\geq$7 per pixel in the \civ\ (1450-1700\A) spectral region.
We discard BAL quasars as cataloged by Scaringi et al. (2009),
since BALs would significantly modify the \civ\ line profile.
12844 quasars meet these criteria. The SDSS spectra are
transformed to the rest frame using the improved redshifts for
SDSS quasars as computed by Hewett \& Wild (2010, hereafter HW10).

To measure the \civ\ line, we first fit the local continuum with a
power-law in two wavelength windows near 1450\A\ and 1695\A. After
subtracting the continuum, we fit the residual spectrum around
\civ\ with two Gaussians. Since the red wing of \civ\ is
contaminated by \heii, only the spectral region of 1450-1580\A\ is
considered in fitting the \civ\ line. The fitting results for most
of the objects are reasonable according to our visual inspection.
However, a small fraction of objects cannot be well fitted by our
automated procedure. To minimize the number of outliers, we
eliminate objects with unacceptable fitting, i.e. $\chi^2/$d.o.f$>1.5$.
This restriction leads to a final working sample of 11268 quasars.

Following Wang11, we use the blueshift and asymmetry index to
measure the deviation of the \civ\ line from an unshifted and
symmetric profile. BAI is defined as the flux ratio of the blue
part to the total profile, where the blue part is the part of
\civ\ line at wavelengths short of 1549.06\A, the rest-frame
wavelength of the \civ\ doublets. The BAI estimation may be
affected by the accuracy of the redshifts which we use to
transform the observed spectra to the rest frame. To reduce the
possible uncertainty of BAI thus introduced, we adopt the
redshifts provided by HW10, which were derived by cross
correlating observed spectra with a carefully-constructed
template. It has been demonstrated by Wang11 that the redshifts of
HW10 do not introduce any significant bias in measuring BAI.

\subsection{Composite-Spectra Construction and Emission-Line
Measurement}\label{sec_ccem}

\begin{table}
\caption{Continuum windows and number of Gaussians}
\label{tab_fit}
\begin{tabular}{lccccccccc}
\hline\hline
Lines & $\lambda_{lo}$(\A) & $\lambda_{hi}$(\A) & Gaussians \\
\hline
\ovi$+$\lyb\ & 1012 & 1055 & 1\\
\oi$+$\siii\ & 1286 & 1321 & 2\\
\cii\ & 1321 & 1353 & 2\\
\siiv$+$\oiv\ & 1353 & 1450 & 3\\
\civ\ & 1450 & 1695 & 3\\
\niv\ & 1695 & 2030 & 1\\
\niii\ & 1690 & 2030 & 1\\
\siii$+$\neiii\ & 1690 & 2030 & 1\\
\aliii\ & 1690 & 2030 & 1\\
\ciii$+$\siiii\ & 1690 & 2030 & 2\\
\hline
\end{tabular}
\tablecomments{$\lambda_{lo}$ and $\lambda_{hi}$ indicate the
center wavelengths of two continuum windows.}
\end{table}

\begin{figure}
\epsscale{1}\plotone{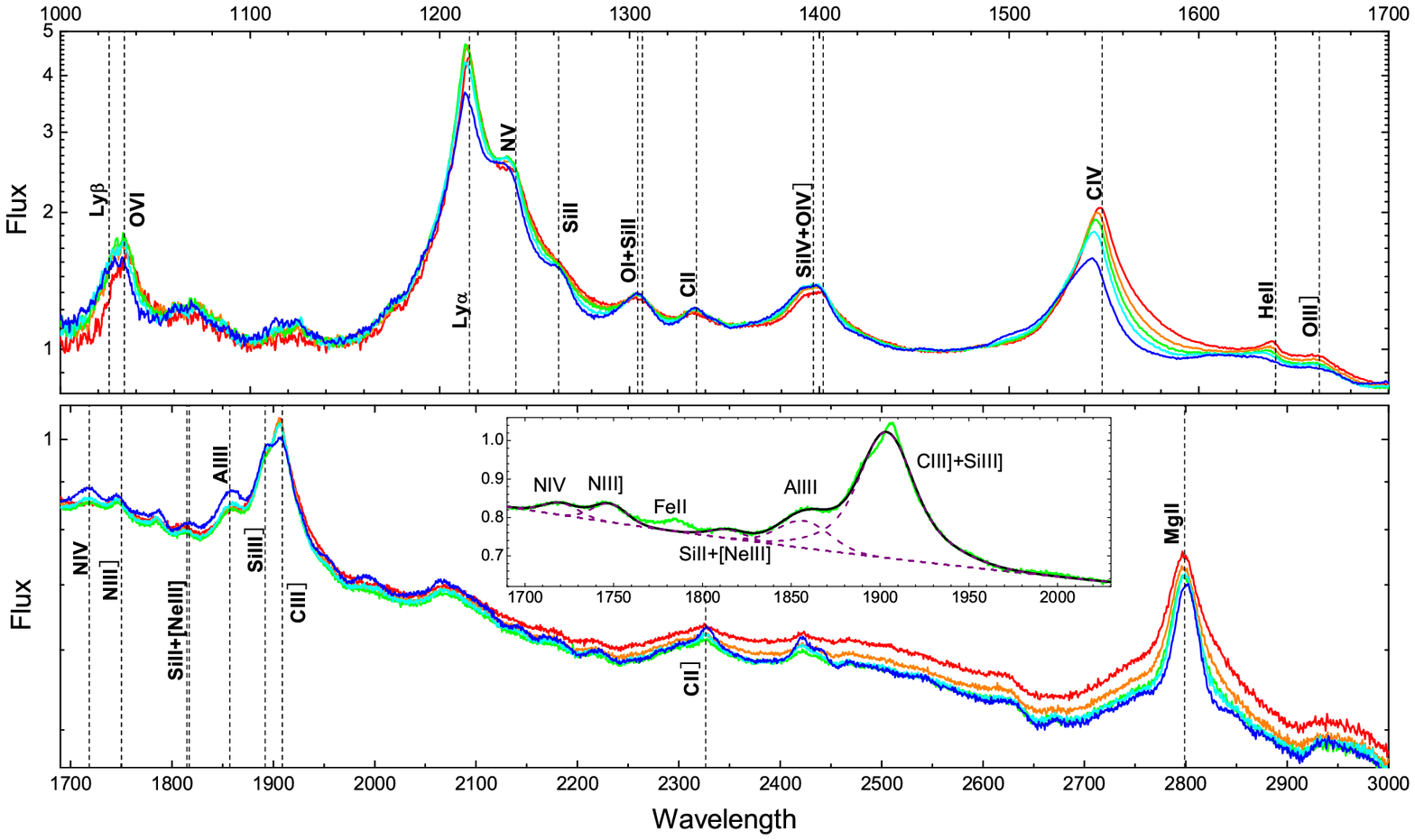}\caption{The composite quasar
spectra, in the order of increasing BAI, are plotted as red,
orange, green, cyan and blue, respectively. Most of the emission
features are labelled. The dashed lines indicate the rest frame
wavelength of the corresponding emission lines. The inset shows a
demonstration of spectral fitting in $1700{\rm \AA}\sim2300{\rm
\AA}$ region for the middle-BAI composite (green line). The black
line shows the best-fit model, the purple dashed lines show the
fitting results for each emission line. The model is described in detail in
Section \ref{sec_ccem}.}\label{fig_cs}
\end{figure}

\begin{figure}
\epsscale{0.7}\plotone{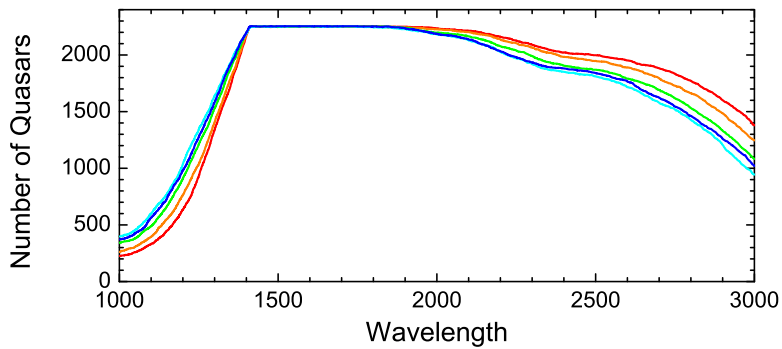}\caption{Number of quasars in each
wavelength bin of the composites as a function of wavelength. The
color code is the same as those in Figure
\ref{fig_cs}.}\label{fig_num}
\end{figure}

With increased S/N, composite spectra are useful for studying weak
lines and the overall properties of quasars by averaging out
object-to-object variations (Vanden Berk et al. 2001, hereafter
VB01; Nagao et al. 2006, hereafter N06). Since we are interested
in the relative intensities of BELs, the composites are
constructed using the arithmetic mean method. We first divide the
quasar sample into five equally-sized subsamples according to
their BAI. For each quasar, we use the HW10 redshift to deredshift
the spectrum. The spectrum is then normalized at 1450\A, rebinned
into the same wavelength bins. In Figure \ref{fig_cs}, we show the
composite spectra covering the wavelength range from 1000 to
3000\A. In our sample, BAI is weakly correlated with redshift and
luminosity, so the number of quasars in each wavelength bin as a
function of wavelength is similar for different composites (Figure
\ref{fig_num}). At least 220 and 940 spectra are combined in the
bins of the composites near 1000\A~and 3000\A, respectively. This
enables detections of weak features in the composite spectra at
high statistical significance. Following VB01 and Laor et al.
(1997), we locate and identify most of significant line features
in the composites.

As \civ\ BAI increases, other high ionization lines, such as
\heii, \nv\ and \ovi, also significantly shift towards shorter
wavelength. Especially, the blue wing of these lines are
prominently enhanced as BAI increases, suggesting that BAI is a
good indicator of outflow strength. While, in our composites, low
ionization lines, such as \oi, \ciii\ and \mgii, don't exhibit
apparent shift with respect to the rest frame. It is
consistent with previous findings (Marziani et al. 1996; Richards
et al. 2002; Shen et al. 2007). These results support that there is no
significant bias in HW10 redshifts and BAI.

We then measure the fluxes of the metal lines (or line complex)
listed in Tab. \ref{tab_fit} from the composites. Generally, we
first fit the continuum with a power-law in two continuum windows.
After subtracting the continuum, we fit the residual line spectrum
with multi-Gaussians model. The central wavelengths of the
continuum windows and the number of Gaussians for each line are
listed in Table \ref{tab_fit}. Note that we only use the spectral region
of 1450-1580\A\ to fit the \civ\ line, as shown in Section \ref{sec_ss}.
We don't measure \nv, \siii, \ctwo\ and \mgii\ (also labelled in Figure \ref{fig_cs}),
because these lines are severely blended with either strong \lya\ or strong \feii\
bump. Due to the uncertainty in the origin of the 1600\A\ bump (see N06), it
is hard to determine the line profile of \heii. We don't investigate \heii\ and \oiii\ too.

The errors of the line fluxes shown in Figure \ref{fig_lr} are
propagated from the uncertainties of the composites. Each
composite is built from more than 2000 quasars, the flux errors
are therefore very small. The errors include none from the
uncertainties in continuum estimate, which is difficult to
quantify. However, the continuum placement is important for weak
line estimates. In the spectral region of 1690-2030\A, several
weak lines are mildly blended. To decompose these lines, we select
spectral regions near 1695\A\ and 2030\A, where the emission line
contribution to the total flux appears to be small, as continuum
windows. We mask the region of 1760-1800\A, where \feii\ emission
is prominent, then simultaneously fit the \niv, \niii,
\siii$+$\neiii\ and \ciii$+$\siiii\ lines. Figure \ref{fig_cs}
shows an example for the spectral decomposition. Our best-fit
model reproduces the composite. \oi$+$\siii\ in small BAI
composites appears to lie on the red wing of the strong line
complex \lya$+$\nv, we therefore adopt local continuums around the
line as continuum windows (see Table \ref{tab_fit}). Nevertheless,
the high continuum around \oi$+$\siii\ at small BAI might be
partly contributed by the broad profile of \oi$+$\siii. If this is
the case, the \oi$+$\siii\ intensity at small BAI is
underestimated and more detailed decomposition is required to
study its dependence on BAI.

\section{BAI and Metallicity Correlation}\label{sec_bmc}

\begin{figure}
\epsscale{1}\plotone{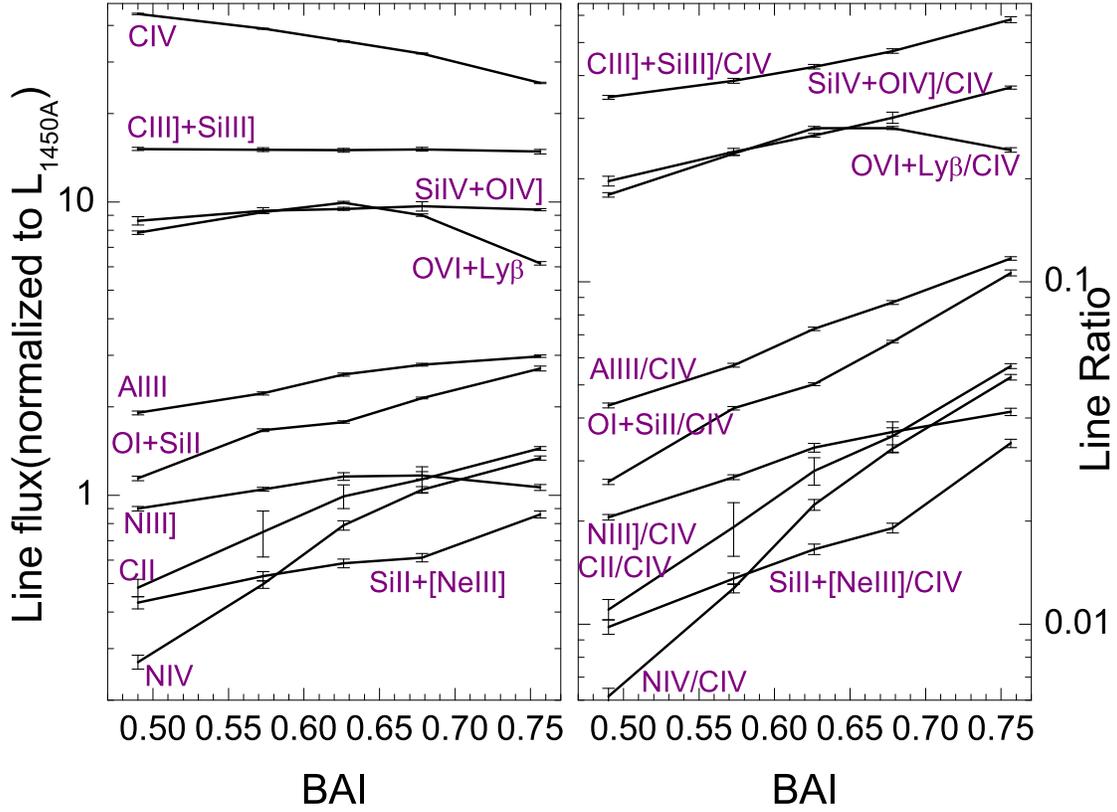}\caption{Left panel: Various line
flux as a function of BAI. Right panels: The flux ratio of other
lines to \civ\ as a function of BAI. All these quantities are
measured from the composites. Note that \ovi$+$\lyb\ are severely
affected by \lya\ forest.}\label{fig_lr}
\end{figure}

The line fluxes as a function of BAI are shown in the left panel
of Figure \ref{fig_lr}. One can see the line intensities are
correlated with BAI, and there seems to exist opposite trends
between strong and weak lines. Strong lines (\civ, \ovi$+$\lyb,
\ciii$+$\siiii\ and \siiv$+$\oiv) either decrease or remain
constant with the increase of BAI, while weak lines (the other
lines shown in Figure \ref{fig_lr}) are dramatically enhanced at
large BAI. The opposite trends also appears to exist in the lines,
which are not measured: the strong line \mgii\ and the weak lines
\siii, \siiii\ and \ctwo\ (see Figure \ref{fig_cs} and Wang11). It
is well known that line equivalent widths decrease with quasar
luminosity, known as the Baldwin effect (Baldwin 1977; Dietrich et
al. 2002). If BAI was correlated with luminosity, one would expect
that line intensities are correlated with BAI. We compute the
median luminosity of each subsample, and find that quasars in the
largest-BAI subsample are, on average, only 45\% more luminous
than those in the smallest-BAI subsample. Apparently, BAI is
weakly related to luminosity\footnote{N06 found a significant
correlation between \civ\ blueshift and luminosity. We note that
they divided the sample into subsamples in a very different way
from us. Most of quasars in their sample are assigned into one or
two subsamples, while our subsamples are equal size. It might be
the most important reason for this difference. } and Baldwin
effect is not important for the trends reported in this paper.

We then show the line ratios of other ions to \civ\ as a function
of BAI in the right panel of Figure \ref{fig_lr}. All line ratios,
except \ovi$+$\lyb/\civ, positively correlate with BAI with a high
statistical significance. Among these ratios, \siiv$+$\oiv/\civ,
\aliii/\civ, \cii/\civ\ and \ciii$+$\siiii/\civ\ have been
demonstrated, by the photoionization model, to significantly
increase with increasing metallicity (e.g. N06). This means that
there exists strong correlation between BAI and metallicity. The
intensities of the strong metal lines depend weakly on metallicity
because of the thermostat effect introduced by the strong cooling
lines (Korista et al. 1998; Ferland 1999). In metal-rich gas,
\civ\ and \ovi \footnote{Note that the measured intensity of
\ovi$+$\lyb\ is severely affected by \lya\ forest, its intrinsic
intensity should be much higher than the measured value.}, the
most important coolants, even tend to decrease with increasing
metallicity because their relative importance as coolants
decreases. While, the weak lines can prominently increase with
metallicity because the number density of corresponding ions
increases and the thermostat effect for them is trivial.
Therefore, the metallicity-BAI correlation can well explain the
trends of the line intensities shown in the left panel of Figure
\ref{fig_lr}. We note that metallicity can not be the only factor
affecting line emission, other factors, such as the ionizing
continuum and the overall properties of BELR gas, also play an
important role.

\begin{figure}
\epsscale{.7}\plotone{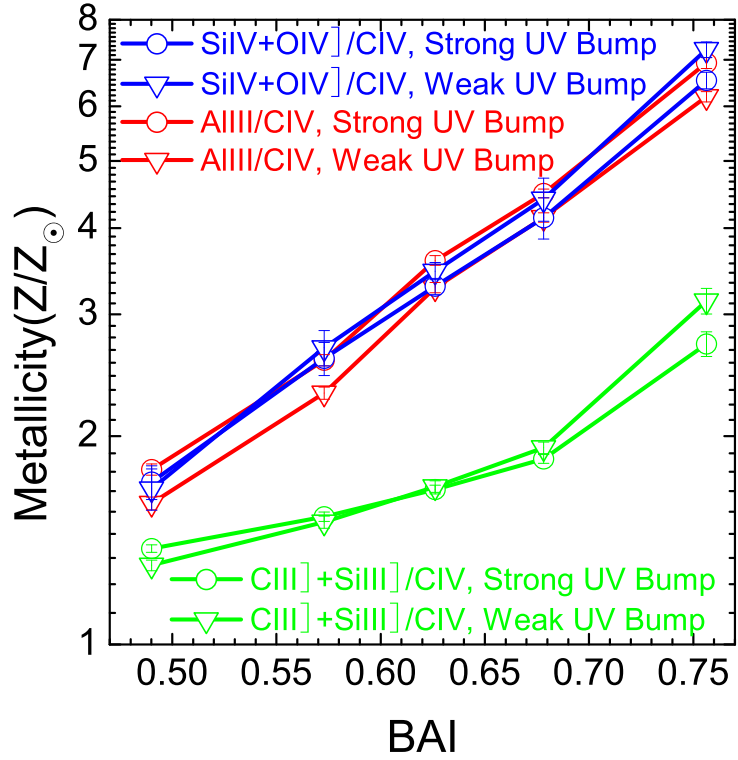}\caption{Metallicity inferred from
different indicators (as indicated in the panel) as a function of
BAI. Please see Nagao et al. (2006) for the details about the
theoretical line ratio-metallicity relationships for different UV
continuum models.}\label{fig_mbai}
\end{figure}

Considerable efforts had been devoted to developing line ratios, such as \nv/\civ\ and \siiv$+$\oiv/\civ,
as metal abundance diagnostics (Hamann et al. 2002; N06).
Here, we adopt \siiv$+$\oiv/\civ\ as a metallicity indicator. Two
theoretical line ratio-metallicity relationships, calculated by
photoionization models using two types of ionizing continuums, are
adopted (N06, Table 10). As shown in Figure \ref{fig_mbai}, the
correlation between the inferred metallicity and BAI is weakly
dependent on the ionizing continuum. When BAI increases from 0.49
to 0.76, the metallicity remarkably increases from 1.7 to
6.9$Z_{\odot}$.

We use this indicator rather than that involving \nv\ because
\siiv$+$\oiv\ is a well isolated feature and the origin of the
\nv\ line might be complex (Wang et al. 2010a). The theoretical
calculation in N06 shows that \cii/\civ, \aliii/\civ\ and
\ciii$+$\siiii/\civ\ are also correlated with metallicity. For
comparison, we also show the metallicities inferred from
\ciii$+$\siiii/\civ\ and \aliii/\civ\ as a function of BAI in
Figure \ref{fig_mbai}. The measured ratios of \cii\ to \civ\ for
the two smallest-BAI subsamples are less than the theoretical
ratios at 0.2$Z_{\odot}$, which is the lowest metallicity adopted
in the theoretical calculation, so we don't calculate the `\cii\
metalliciy'. One can see that the `\aliii\ metallicity' is in good
agreement with the `\siiv$+$\oiv\ metalliciy', implying that the
\aliii\ measurement is reliable. The `\ciii$+$\siiii\ metallicity'
also increases with BAI, although it is apparently lower than the
`\siiv$+$\oiv\ metalliciy'. N06 also found that `\cii\ and
\ciii$+$\siiii\ metallicities' are much lower than `\aliii\ and
\siiv$+$\oiv\ metallicities'. \cii\ and \ciii$+$\siiii\ are
emitted by gas with different ionization degree or gas density
from \civ. This indicates that \cii/\civ\ and \ciii$+$\siiii/\civ\
are sensitive to the BELR properties adopted by photoionization
model, such as the density and spatial distributions of BELR
clouds (see N06 for more details). In contrast, \siiv$+$\oiv/\civ\
is less sensitive to the variation in the BELR structure, and thus
is a more robust metallicity indicator.

\section{Summary and Discussion}\label{secsum}

We measure various line fluxes and the blueshift and asymmetry index (BAI) of the \civ\
line from composite quasar spectra built from the
Sloan Digital Sky Survey. Most of investigated line-flux ratios increase
significantly with BAI. Based on phototionization models, we interpret these dependencies
in terms of stronger outflow in higher gas metallicity environment. This correlation
can also account for the positive trends for weak lines and negative trends for strong lines,
which are observed in the composites. We then use the flux ratio of \siiv$+$\oiv\ to \civ\ as
abundance diagnostics and find that metallicity increases more than 3 times as
BAI increases from 0.49 to 0.76.

We recall that a remarkably similar correlation has been well
established in radiation-driven stellar wind (Massey 2003). High
metallicity leads to large gas opacity, thus large absorption and
acceleration. Recent studies on quasar outflows also favor the
paradigm of radiative acceleration (Wang11). So the
metallicity-outflow correlation can be naturally understood in
terms of the metallicity-acceleration relationship. Another
possibility is that starburst in the galactic center can produce
abundant metals and drive the gases to fuel the black hole (Wang
et al. 2010b). Because outflows are boosted at high accretion
rates, the BAI$-$metallicity relationship can also be produced.
The two mechanisms may both work. Although it is difficult to
disentangle the predominant physical mechanism from the current
data, the first mechanism is apparently more direct.
Whichever mechanism dominates, star formation occurred in the host galaxies,
which produces metals, has an important impact on the driving of quasar outflow.

The quasar metallicity may dramatically affect the host galaxy
properties via feedback. On one hand, outflows from the quasar
could exert enough force on the ISM to remove some of the material
from the galaxy, consequently inhibiting the growth of the galaxy.
This process would be more effective at higher metallicity
according to our results. On the other hand, metal-rich outflows
and ejected ISM may chemically enrich the IGM as the BELR
metallicity is much higher than the solar value (Hamann et al.
2002). The enrichment makes the radiative cooling of the IGM more
efficient, and further enhances star formation in galaxies and the
growth of their central black holes, in contrast to the momentum
feedback process that blows away the ISM. The enrichment rate is
proportional to both the mass loss rate of outflows and
metallicity, and therefore increases with metallicity in a manner
stronger than a linear relationship. This feedback process is
particularly important at high redshifts because 1) the intense
star formation at high redshifts may form a substantially
metal-rich environment in the galactic nuclei; 2) at high
redshifts, IGM is very metal poor and its cooling is less
effective.

\acknowledgements We thank the referee for constructive comments that significantly
improved the paper. This work is supported by NSFC (11073017, 11033007, 10973013, 10973012), 973
program (2007CB815405, 2009CB824800). Funding
for the SDSS and SDSS-II has been provided by the Alfred P. Sloan
Foundation, the Participating Institutions, the National Science
Foundation, the U.S. Department of Energy, the National
Aeronautics and Space Administration, the Japanese Monbukagakusho,
the Max Planck Society, and the Higher Education Funding Council
for England. The SDSS Web site is http://www.sdss.org/.

%\end{CJK*}

\begin{thebibliography}{}

\bibitem[Baldwin(1977)]{1977ApJ...214..679B} Baldwin, J.~A.\ 1977, \apj,
214, 679

\bibitem[Crenshaw et
al.(2003)]{2003ARA&A..41..117C} Crenshaw, D.~M., Kraemer, S.~B.,
\& George, I.~M.\ 2003, \araa, 41, 117

\bibitem[Dietrich et al.(2002)]{2002ApJ...581..912D} Dietrich, M., Hamann,
F., Shields, J.~C., et al.\ 2002, \apj, 581, 912

\bibitem[Di Matteo et al.(2005)]{2005Natur.433..604D} Di Matteo, T.,
Springel, V., \& Hernquist, L.\ 2005, \nat, 433, 604

\bibitem[Fabian(1999)]{1999MNRAS.308L..39F} Fabian, A.~C.\ 1999, \mnras,
308, L39

\bibitem[Fan et al.(2009)]{2009ApJ...690.1006F} Fan, L.~L., Wang, H.~Y.,
Wang, T., Wang, J., Dong, X., Zhang, K., \& Cheng, F.\ 2009, \apj,
690, 1006

\bibitem[Ferland(1999)]{1999ASPC..162..147F} Ferland, G.\ 1999, Quasars and
Cosmology, 162, 147

\bibitem[Friaca
\& Terlevich(1998)]{1998MNRAS.298..399F} Friaca, A.~C.~S., \&
Terlevich, R.~J.\ 1998, \mnras, 298, 399

\bibitem[Ganguly et al.(2007)]{2007ApJ...665..990G} Ganguly, R.,
Brotherton, M.~S., Cales, S., et al.\ 2007, \apj, 665, 990

\bibitem[Hamann et al.(2002)]{2002ApJ...564..592H} Hamann, F., Korista,
K.~T., Ferland, G.~J., Warner, C., \& Baldwin, J.\ 2002, \apj,
564, 592

\bibitem[Hewett
\& Wild(2010)]{2010MNRAS.405.2302H} Hewett, P.~C., \& Wild, V.\
2010, \mnras, 405, 2302

\bibitem[Korista et al.(1998)]{1998ApJ...507...24K} Korista, K., Baldwin,
J., \& Ferland, G.\ 1998, \apj, 507, 24

\bibitem[Laor et al.(1997)]{1997ApJ...489..656L} Laor, A., Jannuzi, B.~T.,
Green, R.~F., \& Boroson, T.~A.\ 1997, \apj, 489, 656

\bibitem[Leighly
\& Moore(2004)]{2004ApJ...611..107L} Leighly, K.~M., \& Moore,
J.~R.\ 2004, \apj, 611, 107

\bibitem[Marziani et al.(1996)]{1996ApJS..104...37M} Marziani, P.,
Sulentic, J.~W., Dultzin-Hacyan, D., Calvani, M., \& Moles, M.\
1996, \apjs, 104, 37

\bibitem[Massey(2003)]{2003ARA&A..41...15M} Massey, P.\ 2003, \araa, 41, 15

\bibitem[Nagao et
al.(2006)]{2006A&A...447..157N} Nagao, T., Marconi, A., \&
Maiolino, R.\ 2006, \aap, 447, 157

\bibitem[Richards et al.(2002)]{2002AJ....124....1R} Richards, G.~T.,
Vanden Berk, D.~E., Reichard, T.~A., Hall, P.~B., Schneider,
D.~P., SubbaRao, M., Thakar, A.~R., \& York, D.~G.\ 2002, \aj,
124, 1

\bibitem[Richards et al.(2011)]{2011AJ....141..167R} Richards, G.~T., et
al.\ 2011, \aj, 141, 167

\bibitem[Scaringi et al.(2009)]{2009MNRAS.399.2231S} Scaringi, S., Cottis,
C.~E., Knigge, C., \& Goad, M.~R.\ 2009, \mnras, 399, 2231

\bibitem[Schneider et al.(2007)]{2007AJ....134..102S} Schneider, D.~P., et
al.\ 2007, \aj, 134, 102

\bibitem[Shen et al.(2007)]{2007AJ....133.2222S} Shen, Y., Strauss, M.~A.,
Oguri, M., et al.\ 2007, \aj, 133, 2222

\bibitem[Silk
\& Rees(1998)]{1998A&A...331L...1S} Silk, J., \& Rees, M.~J.\ 1998, \aap, 331, L1

\bibitem[Vanden Berk et al.(2001)]{2001AJ....122..549V} Vanden Berk, D.~E.,
Richards, G.~T., Bauer, A., et al.\ 2001, \aj, 122, 549

\bibitem[Wang et al.(2010a)]{2010ApJ...710...78W} Wang, H., Wang, T., Yuan,
W., et al.\ 2010a, \apj, 710, 78

\bibitem[Wang et al.(2010b)]{2010ApJ...719L.148W} Wang, J.-M., Yan, C.-S.,
Gao, H.-Q., et al.\ 2010b, \apjl, 719, L148

\bibitem[Wang et al.(2011)]{2011ApJ...738...85W} Wang, H., Wang, T., Zhou,
H., et al.\ 2011, \apj, 738, 85

\bibitem[Wang et al.(2009)]{2009ApJ...702..851W} Wang, T., Zhou, H., Yuan,
W., Lu, H.~L., Dong, X., \& Shan, H.\ 2009, \apj, 702, 851

\bibitem[Zhang et al.(2010)]{2010ApJ...714..367Z} Zhang, S., Wang, T.-G.,
Wang, H., Zhou, H., Dong, X.-B., \& Wang, J.-G.\ 2010, \apj, 714,
367
\end{thebibliography}
\end{document}